\documentclass{llncs}
\usepackage{algorithm}
\usepackage{algorithmic}

\usepackage{amsmath}
\usepackage{amsfonts}
\usepackage{epsfig}

\newcommand{\card}[1]{\left| #1 \right|}
\newcommand{\reals}{\mathbb{R}}
\newcommand\alld{\texttt{all\-diff\-erent}}
\newcommand\softad{\texttt{soft\_all\-diff\-erent}}
\renewenvironment{proof}{\noindent{\bf Proof.}}{\mbox{}\hfill$\Box$\\}

\begin{document}

\mainmatter

\title{A Hyper-Arc Consistency Algorithm for the \\
Soft Alldifferent Constraint}


\author{Willem Jan van Hoeve\inst{}}

\institute{CWI, P.O. Box 94079, 1090 GB Amsterdam, The Netherlands\\
\email{W.J.van.Hoeve@cwi.nl}\\
\texttt{http://homepages.cwi.nl/\~{ }wjvh/}}

\maketitle
\begin{abstract}
This paper presents an algorithm that achieves hyper-arc consistency 
for the soft \alld\ constraint. To this end, we prove and exploit the
equivalence with a minimum-cost flow problem. Consistency of the 
constraint can be checked in $O(nm)$ time, and hyper-arc consistency 
is achieved in $O(m)$ time, where $n$ is the number of variables 
involved and $m$ is the sum of the cardinalities of the domains. 
It improves a previous method that did not ensure hyper-arc consistency.
\end{abstract}

\section{Introduction}
If a constraint satisfaction problem (CSP) is over-constrained,
i.e. has no solution satisfying all constraints, it is natural to
allow certain constraints, the soft constraints, to be violated and
search for solutions that violate as few soft constraints as
possible. Constraints that are not decided to be soft are hard
constraints, and should always be satisfied.

Several methods have been proposed to handle over-constrained
CSPs, see for instance \cite{freuder92,bmr97,schiex03}. In this 
paper, we follow the scheme proposed by R\'egin, Petit, Bessi\`ere and 
Puget \cite{regin2000over}, that is particularly useful for non-binary 
constraints. The idea is as follows. A cost function is
assigned to each soft constraint, measuring the violation. Then the
soft CSP is transformed into a constraint optimization problem (COP),
where all constraints are hard, and the (weighted) sum of cost
functions is minimized. This approach allows one to use specialized
filtering algorithms for soft constraints, as shown by
Petit, R\'egin and Bessi\`ere \cite{petit01}.

For the soft \alld\ constraint, an algorithm is presented in 
\cite{petit01} that removes inconsistent values in $O(m^2 n \sqrt{n})$ 
time, where $n$ is the number of variables and $m$ the sum of the
cardinalities of their domains. However, that algorithm does not
ensure hyper-arc consistency. In this paper, we propose an algorithm 
that does ensure hyper-arc consistency and runs in $O(nm)$ time. 
In principle, we consider the soft \alld\ constraint as a minimum-cost
flow problem in a particular graph. Checking the consistency can then 
be done in $O(nm)$ time. Thereafter, domain values are checked for 
consistency by an efficient shortest path computation, which takes in 
total $O(m)$ time.

The outline of the paper is as follows. Section~\ref{sc:prel} presents
definitions related to constraint satisfaction problems. 
Section~\ref{sc:graph} shows a graph-theoretic analysis of the soft
\alld\ constraint, using flow theory. In Section~\ref{sc:alg} the
filtering algorithm is presented. We conclude with a discussion in
Section~\ref{sc:concl}.

\section{Background}\label{sc:prel}
We assume familiarity with the basic concepts of constraint
programming. For a thorough explanation of constraint programming,
see~\cite{apt2003}.

A constraint satisfaction problem (CSP) consists of a finite set of
variables ${\cal V} = \{ v_1, \dots , v_r \}$ with finite domains 
${\cal D}= \{ D_1, \dots, D_r \}$ such that $v_i \in D_i$ for all $i$, 
together with a finite set of constraints ${\cal C}$, each on a subset
of ${\cal V}$. 
A constraint $C \in {\cal C}$ is defined as a subset of the Cartesian 
product of the domains of the variables that are in $C$. 
A tuple $(d_1, \dots, d_r) \in D_1 \times \dots \times D_r$ is a
solution to a CSP if for every constraint $C \in {\cal C}$ on the 
variables $v_{i_1}, \dots , v_{i_k}$ we have 
$(d_{i_1}, \dots, d_{i_k}) \in C$. A constraint optimization problem
(COP) is a CSP together with an objective function to be optimized. A
solution to a COP is a solution to the corresponding CSP, that has an
optimal objective function value.

\begin{definition}[Hyper-arc consistency]\label{def:hac}
A constraint $C$ on the variables $x_1$, $\dots, x_k$ is called 
hyper-arc consistent if for each variable $x_i$ and value
$d_i \in D_i$, there exist values $d_1,\dots, d_{i-1}, d_{i+1}, \dots, d_k$ 
in $D_1, \dots, D_{i-1},D_{i+1}, \dots, D_k$, such that $(d_1, \dots , d_k) 
\in~C$. 
\end{definition}

\begin{definition}[Consistent CSP]
A CSP is hyper-arc consistent if all its constraints are hyper-arc
consistent. A CSP is inconsistent if it has no solution. Similarly for
a COP.
\end{definition}

\begin{definition}[Pairwise difference]\label{def:alldiff}
Let $x_1, \dots, x_n$ be variables with respective finite domains
$D_1, \dots, D_n$. Then 
\begin{displaymath}
\begin{array}{l}
\textup{\alld}(x_1, \dots, x_n) = 
\{ (d_1, \dots, d_n) \mid d_i \in D_i, d_j \neq d_k \; {\rm for } \; j \neq k \}.
\end{array}
\end{displaymath}
\end{definition}

In \cite{petit01}, two different measures of violation for a soft
constraint are presented. The first is the minimum number of variables 
that need to change their value in order to satisfy the constraint. 
For this measure, applied to the \alld\ constraint, \cite{petit01} also 
contains a hyper-arc consistency algorithm. The second 
measure is the number of violated constraints in the binary 
decomposition of the constraint, if this decomposition exists. 
For the \alld\ constraint, such a decomposition does exist, namely
$x_i \neq x_j$ for $i \in \{1, \dots, n-1\}, j \in \{i+1, \dots, n\}$.
We follow this second, more refined, measure, and present it in 
terms of the soft \alld\ constraint. For \alld$(x_1, \dots, x_n)$, let
the cost of violation be defined as
\begin{equation}\label{eq:objective}
{\rm violation}(x_1, \dots, x_n) = \card{ \{ (i,j) \mid x_i = x_j, 
\;{\rm for}\; i < j \} }.
\end{equation}
\begin{definition}[Soft pairwise difference]\label{def:softad}
Let $x_1, \dots, x_n, z$ be variables with respective finite domains
$D_1, \dots, D_n, D_z$. Then 
\begin{displaymath}
\begin{array}{l}
\textup{\softad}(x_1, \dots, x_n, z) = \\
\{ (d_1, \dots, d_n, \tilde{d}) \mid d_i \in D_i, \tilde{d} \in D_z, 
{\rm violation}(d_1, \dots, d_n) \leq \tilde{d} \}.
\end{array}
\end{displaymath}
\end{definition}
The variable $z$ in Definition~\ref{def:softad} will serve as a 
so-called cost variable, which will be minimized during the solution 
process. This means that admissible tuples in 
Definition~\ref{def:softad} are those instantiations of variables, 
such that the number of violated dis-equality constraints 
$d_i \neq d_j$ is not more than that of the currently best found 
solution, represented by $\max D_z$. At the same time, $\min D_z$
should not be less than the currently lowest possible value of 
${\rm violation}(x_1, \dots, x_n)$.

An over-constrained CSP with an \alld\ constraint is transformed
into a COP by introducing $z$, replacing \alld\ with \softad, and
minimizing $z$. This is illustrated in the following example.
\begin{example}\label{ex:over}
Consider the following over-constrained CSP
\begin{displaymath}
\begin{array}{l}
x_1 \in \{a,b\}, x_2 \in \{a,b\}, x_3 \in \{a,b\}, x_4 \in \{b,c\}, \\
\textup{\alld}(x_1, x_2, x_3, x_4).
\end{array}
\end{displaymath}
We transform this CSP into 
\begin{displaymath}
\begin{array}{l}
z \in \{0, \dots, 6\},\\
x_1 \in \{a,b\}, x_2 \in \{a,b\}, x_3 \in \{a,b\}, x_4 \in \{b,c\}, \\
\textup{\softad}(x_1, x_2, x_3, x_4, z),\\
\textup{\tt minimize} \; z.
\end{array}
\end{displaymath}
This COP is not hyper-arc consistent, as there is no support for $z<1$.
If we remove $0$ from $D_z$, the COP is hyper-arc consistent, because 
there are at most 6 simultaneously violated dis-equalities.
Suppose now that during the search for a solution, 
we have found the tuple $(x_1, x_2, x_3, x_4, z) = (a,a,b,c,1)$, that has 
one violated dis-equality. Then $z \in \{1\}$ in the remaining search.
As the assignment $x_4 = b$ always leads to a solution with $z \geq 2$,
$b$ can be removed from $D_4$. The resulting COP is hyper-arc consistent 
again.
\end{example}

One should take into account that a simplified CSP is considered in 
Example~\ref{ex:over}. In general, a CSP can consist of many more 
constraints, and also more cost-variables that together with $z$ form
an objective function to be minimized.

Throughout this paper, let $m = \sum_{i \in \{1, \dots, n \}} \card{D_i}$
for variables $x_1, \dots, x_n$.

\section{Graph-Theoretic Analysis}\label{sc:graph}
A directed graph is a pair $G = (V,A)$ where $V$ is a finite set of 
vertices $V$ and $A$ is a family\footnote{A family is a set in which 
elements may occur more than once.} of ordered pairs from $V$, called arcs. 
A pair occurring more than once in $A$ is called a multiple arc.
For $v \in V$, let $\delta^{\rm in}(v)$ and $\delta^{\rm out}(v)$ denote 
the family of arcs entering and leaving $v$ respectively.

A (directed) walk in $G$ is a sequence $P = v_0, a_1, v_1, \dots, a_k,
v_k$ where $k \geq 0$, $v_0, v_1, \dots, v_k \in V$, $a_1, \dots, a_k
\in A$ and $a_i = (v_{i-1}, v_i)$ for $i = 1, \dots, k$. If there is
no confusion, $P$ may be denoted as $P = v_0, v_1, \dots, v_k$. A
(directed) walk is called a (directed) path if  $v_0, \dots, v_k$ are
distinct. A closed (directed) walk, i.e. $v_0 = v_k$, is called a
(directed) circuit if $v_1, \dots, v_k$ are distinct.

\subsection{Minimum-cost flow problem}\label{ssc:flow}
First, we introduce the concept of a flow, following 
Schrijver~\cite[pp. 148--150]{lex2003}.

Let $G = (V,A)$ be a directed graph and let $s,t \in V$.
A function $f: A \rightarrow \reals$ is called a flow from $s$ to $t$, or 
an $s-t$ flow, if
\begin{equation}\label{eq:flow}
\begin{array}{rll}
(i) \; & f(a) \geq 0 & \textrm{for each } a \in A, \\
(ii) \; & f(\delta^{\rm out}(v)) = f(\delta^{\rm in}(v)) & 
\textrm{for each } v \in V \setminus \{s,t\},
\end{array}
\end{equation}
where $f(S) = \sum_{a \in S} f(a)$ for all $S \subseteq A$.
Property (\ref{eq:flow})$(ii)$ ensures flow conservation, i.e. for
a vertex $v \neq s,t$, the amount of flow entering $v$ is equal to the 
amount of flow leaving $v$. 

The value of an $s-t$ flow $f$ is defined as
\begin{displaymath}
{\rm value}(f) = f(\delta^{\rm out}(s)) - f(\delta^{\rm in}(s)).
\end{displaymath}
In other words, the value of a flow is the net amount of flow leaving $s$, 
which can be shown to be equal to the net amount of flow entering $t$.

When we study flows we typically endow capacity constraints, via 
a ``capacity'' function $c:A \rightarrow \reals_+$. We say that a
flow $f$ is under $c$ if $f(a) \leq c(a)$ for each $a \in A$. A feasible
flow is a flow under $c$.

We also assign costs to flows via a ``cost'' function 
$w:A \rightarrow \reals_+$. Doing so the cost of a flow $f$ is defined as 
\begin{displaymath}
{\rm cost}(f) = \sum_{a \in A} w(a) f(a).
\end{displaymath}

A minimum-cost flow is an $s-t$ flow under $c$ of maximum value and 
minimum cost. The minimum-cost flow problem is the problem of finding 
such a minimum-cost flow.

A minimum-cost flow can be computed using an algorithm originally due 
to Ford and Fulkerson~\cite{ford1958} (we follow the description given
by Schrijver \cite[pp. 183--185]{lex2003}). It consists of successively
finding shortest (with respect to the cost function) $s-t$ paths in the 
so-called residual graph, while maintaining an optimal flow.

Define the residual graph $G_f = (V,A_f)$ of $f$ (with respect to $c$),
where 
\begin{displaymath}
A_f = \{a \mid a \in A, f(a) < c(a)\} \cup 
\{a^{-1} \mid a \in A, f(a) > 0 \}.
\end{displaymath}
Here $a^{-1} = (v,u)$ if $a = (u,v)$. We extend $w$ to 
$A^{-1} = \{a^{-1} \mid a \in A\}$ by defining
\begin{displaymath}
w(a^{-1}) = - w(a)
\end{displaymath}
for each $a \in A$.

Any directed path $P$ in $G_f$ gives an undirected path in $G = (V,A)$. 
We define $\chi^P \in \reals^A$ by
\begin{displaymath}
\chi^P(a) = \left\{
  \begin{array}{rl}
  1 & \textrm{if $P$ traverses $a$},\\
 -1 & \textrm{if $P$ traverses $a^{-1}$},\\
  0 & \textrm{if $P$ traverses neither $a$ nor $a^{-1}$},\\
  \end{array} 
\right.
\end{displaymath}
for $a \in A$. Define the cost of a path $P$ as 
${\rm cost}(P) = \sum_{a \in P} w(a)$.

Call a feasible flow extreme when it has minimum cost among all feasible 
flows with the same value. Then the following holds 
(cf. \cite[Theorem 12.3 and 12.4]{lex2003}). Let ${\bf 0}$ denote the 
all-zero vector of appropriate size.
\begin{theorem}\label{thm:circuit}
A flow $f$ is extreme if and only if each directed circuit of $G_f$ 
has nonnegative cost.
\end{theorem}
\begin{theorem}\label{thm:extreme}
Let $f$ be an extreme flow in $G = (V,A)$. Let $P$ be a minimum-cost 
$s-t$ path in $G_f$, for some $s,t \in V$, and let $\varepsilon > 0$ be 
such that $f^\prime = f + \varepsilon \chi^P$ satisfies 
${\bf 0} \leq f^\prime \leq c$. 
Then $f^\prime$ is an extreme flow again.
\end{theorem}

In fact, for $f, P, \varepsilon$ and $f^\prime$ in Theorem~\ref{thm:extreme} 
holds
\begin{displaymath}
\begin{array}{l}
{\rm value}(f^{\prime}) = {\rm value}(f)+\varepsilon,\\
{\rm cost}(f^{\prime}) = {\rm cost}(f)+\varepsilon \cdot {\rm cost}(P).
\end{array}
\end{displaymath}

\begin{algorithm}[t]
\begin{algorithmic}
  \STATE{set $f = {\bf 0}$}
  \WHILE{termination criterion not satisfied}
    \STATE{compute minimum-cost $s-t$ path $P$ in $G_f$}
    \IF{no $s-t$ path in $G_f$ exists}
      \STATE{terminate}
    \ELSE{}
      \STATE{set $\varepsilon$ maximal, such that 
        ${\bf 0} \leq f + \varepsilon \chi^P \leq c$}
      \STATE{reset $f = f + \varepsilon \chi^P$}
    \ENDIF
  \ENDWHILE
\end{algorithmic}
\caption{Minimum-cost $s-t$ flow} \label{alg:mincostflow}
\end{algorithm}

This means that we can find a minimum-cost $s-t$ flow in $G$ by 
successively computing minimum-cost $s-t$ paths in $G_f$. Along such
a path we increase the amount of flow to the maximum possible value
$\varepsilon$. By Theorem~\ref{thm:extreme}, the last flow (of maximum 
value) we obtain must be extreme, and hence optimal. This is presented 
as Algorithm~\ref{alg:mincostflow}. Note that the cost of minimum-cost 
$s-t$ paths in $G_f$ is bounded, because there are no directed circuits 
of negative cost in $G_f$. For rational capacities, 
Algorithm~\ref{alg:mincostflow} terminates with a feasible $s-t$ flow 
of maximum value and minimum cost.
Although faster algorithms exist for general minimum-cost flow 
problems, Algorithm~\ref{alg:mincostflow} suffices when applied to
our problem. This is because in our particular graph 
Algorithm~\ref{alg:mincostflow} is faster than the algorithms for
general minimum-cost flow problems.

\subsection{From \softad\ to minimum-cost flow}\label{ssc:softflow}
We transform the problem of finding a solution to the \softad\
constraint into a minimum-cost flow problem.

Construct the directed graph $G = (V,A)$ with
\begin{displaymath}
V = \{s,t\} \cup X \cup D_X
\end{displaymath}
and
\begin{displaymath}
A = A_X \cup A_s \cup A_t
\end{displaymath}
where
\begin{displaymath}
\begin{array}{l}
X = \{x_1, \dots, x_n\}, \\
D_X = \bigcup_{i \in \{1, \dots, n\}} D_i, \\
\end{array}
\end{displaymath}
and
\begin{displaymath}
\begin{array}{rl}
A_X =& \{(x_i, d) \mid d \in D_i \}, \\
A_s =& \{ (s, x_i) \mid i \in \{1, \dots, n\} \}, \\
A_t =& \{ (d, t) \mid d \in D_i, i \in \{1, \dots, n\} \}.
\end{array}
\end{displaymath}
Note that $A_t$ contains parallel arcs if two or more variables
share a domain value. If there are $k$ parallel arcs $(d,t)$ between
some $d \in D_X$ and $t$, we distinguish them by numbering the arcs as
$(d,t)_0, (d,t)_1, \dots, (d,t)_{k-1}$ in a fixed but arbitrary way.

To each arc $a \in A$, we assign a capacity $c(a) = 1$ and a cost 
$w(a)$. If $a \in A_s \cup A_X$, then $w(a) = 0$. If $a \in A_t$, 
so $a = (d, t)_i$ for some $d \in D_X$ and integer $i$, the value 
of $w(a) = i$.

\begin{figure}[t]
\begin{center}
\epsfig{figure=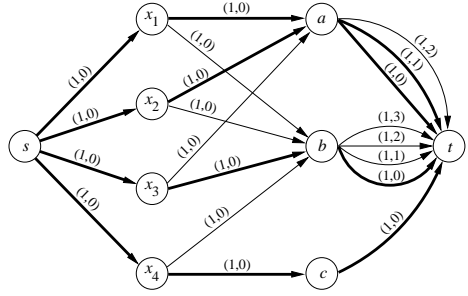,width=.5\textwidth} 
\end{center}
\caption{Graph $G$ for the \softad\ constraint of
  Example~\ref{ex:over}. For each arc $a$, $(c(a), w(a))$ is given. Bold
arcs indicate an optimal $s-t$ flow with cost 1.} \label{fig:network}
\end{figure}

In Figure~\ref{fig:network}, the graph $G$ for the \softad\ constraint 
in Example~\ref{ex:over} is depicted. For each arc $a$, $(c(a),
w(a))$ is given. 

\begin{theorem}\label{thm:flow}
An integer flow $f$ that is a solution to the minimum-cost flow
problem in $G$ corresponds to an instantiation of variables 
$x_1, \dots, x_n$ in \textup{\softad}$(x_1, \dots, x_n, z)$, minimizing 
${\rm violation}(x_1, \dots, x_n)$.
\end{theorem}
\begin{proof}
For an integer flow $f$ in $G$, $f(a) = 1$ if arc $a$ is used, and
$f(a) = 0$ otherwise. An arc $a = (x_i, d) \in A_X$ with $f(a) = 1$
corresponds to the instantiation $x_i = d$. By construction, every
solution $f$ to the minimum-cost flow problem in $G$ has 
${\rm value}(f) = n$. Thus a solution corresponds to assigning a value
to each variable $x_i$, $i \in \{1, \dots, n\}$.

The cost function $w(a_i) = i$ for $k$ parallel arcs $a_0, \dots,
a_{k-1} \in A_t$ corresponds to counting the number of violations
caused by assigning $i+1$ variables to a particular value.
Namely, for these parallel arcs, a minimum-cost $s-t$ path in $G_f$ 
uses the arc with lowest cost first. Using arc $a_i$ (the $(i+1)$st 
arc) causes a ``violation'' with the $i$ previously used arcs. Thus, 
for a feasible flow $f$, which corresponds to an assignment of 
$x_1, \dots, x_n$, $\sum_{a \in A} w(a)f(a)$ measures exactly 
${\rm violation}(x_1, \dots, x_n)$. Hence, a minimum-cost flow 
minimizes ${\rm violation}(x_1, \dots, x_n)$.
\end{proof}

Consider again the graph $G$ in Figure~\ref{fig:network}. A bold arc 
$a$ in $G$ denotes $f(a) = 1$. This particular flow $f$ has 
value$(f) = 4$ and cost$(f) = 1$. Indeed, the only violation is 
$x_1 = a = x_2$.

Next we describe the behaviour of Algorithm~\ref{alg:mincostflow} 
to compute a minimum-cost flow in $G$. We need to compute 
a sequence of minimum-cost $s-t$ paths in $G_f$, maintaining extreme
intermediate flows. Note that along each minimum-cost $s-t$ path in 
$G_f$ we can increase the flow by a maximum of $\varepsilon = 1$. Hence 
all extreme flows in $G$ are integer. By construction, 
there are exactly $n$ such paths, each containing one arc in $A_s$
(in fact, the paths may as well be computed starting from the vertices 
$x_i$ instead of $s$, using only arcs in $A_X$ and $A_t$). 
Further, each minimum-cost $s-t$ path contains exactly one arc in 
$A_t$. Namely, consider a minimum-cost path $P$ using multiple arcs in 
$A_t$. Then $P$ consists of an $s-t$ path with one arc in $A_s$, 
followed by a $t-t$ path. 
If the $t-t$ path has cost 0, we may omit this part, and use 
only the $s-t$ path with one arc in $A_s$. If the $t-t$ path, which 
is a circuit, has negative cost, it contradicts 
Theorem~\ref{thm:circuit}. Effectively, it means that the $t-t$ path
could have been used to improve the preceding intermediate solution, 
thus contradicting the extremity of that solution.
To conclude, the minimum-cost paths we need to compute use exactly one
arc in $A_s$ and one arc in $A_t$. It follows that these paths can be 
computed in $O(m)$ time, and the total time complexity for finding a 
maximum flow of minimum cost in $G$ is $O(nm)$. Hence it follows, by
Theorem~\ref{thm:flow}, that consistency of the \softad\ constraint
can be checked in $O(nm)$ time.

\section{The Filtering Algorithm}\label{sc:alg}
The following theorem identifies hyper-arc consistent domain
values for the \softad\ constraint. For an arc $a$ of $G$, let $G^a$
arise from $G$ by enforcing $f(a) = 1$ for every flow $f$ in $G$.
\begin{theorem}\label{thm:hac}
The constraint \textup{\softad}$(x_1, \dots, x_n, z)$ is hyper-arc
consistent if and only if 
\begin{itemize}
\item[$(i)$] for all all arcs $a \in A_X$ a minimum-cost flow of 
maximum value in $G^a$ has cost at most $\max{D_z}$,
\item[$(ii)$] all values in $D_z$ are not smaller than the cost of a 
minimum-cost flow of maximum value in $G$.
\end{itemize}
\end{theorem}
\begin{proof}
Enforcing $f(a) = 1$ for arc $a = (x_i, d)$ corresponds to assigning
$x_i = d$. The result follows from Definition~\ref{def:hac} and 
Theorem~\ref{thm:flow}. Namely, property $(i)$ checks consistency for
all domain values in $D_1, \dots, D_n$. Property $(ii)$ checks 
consistency of the domain values of $D_z$.
\end{proof}

\begin{algorithm}[t]
\caption{Naive hyper-arc consistency}
\label{alg:hac}
\begin{algorithmic}
  \STATE{set ${\rm minimum} = \infty$}
  \FOR{$x_i \in X$}
  \FOR{$d \in D_i$}
  \STATE{compute minimum-cost $s-t$ flow $f$ in $G^{a}$ where $a = (x_i, d)$}
  \IF{cost$(f) > \max{D_z}$}
  \STATE{remove $d$ from $D_i$}
  \ENDIF
  \IF{cost$(f) < {\rm minimum}$}
  \STATE{set ${\rm minimum} = {\rm cost}(f)$}
  \ENDIF
  \ENDFOR
  \ENDFOR
  \IF{$\min{D_z} < {\rm minimum}$}
  \STATE{set $\min{D_z} = {\rm minimum}$}
  \ENDIF  
\end{algorithmic}
\end{algorithm}

Using Theorem~\ref{thm:hac}, we can construct an algorithm that
enforces hyper-arc consistency for the \softad\ constraint, presented
as Algorithm~\ref{alg:hac}. For all variables $x_i \in X$, the algorithm 
scans all domain values $d \in D_i$, and checks whether there exists a 
minimum-cost $s-t$ flow in $G^a$, where $a = (x_i,d)$, of maximum value 
with cost at most $\max{D_z}$. If such a flow does not exist,
then, by Theorem~\ref{thm:hac}, $d$ is 
removed from $D_i$. Finally, we remove all values of $D_z$ which are 
smaller than the cost of a minimum-cost flow in $G$. The time complexity 
of Algorithm~\ref{alg:hac} is $O(m^2 n)$. 

We can construct a more efficient filtering algorithm, however. It is
presented as Algorithm~\ref{alg:hac2}, and makes use of the following
theorem. We follow the notation introduced in Section~\ref{ssc:flow}.
\begin{theorem}\label{thm:path}
Let $f$ be an extreme flow of maximum value in $G$. Let $a = (x_i,
d) \in A_X$ and $P$ a minimum-cost $d-x_i$ path in $G_f$. Let
$f^\star$ be an extreme flow of maximum value in $G^a$. Then 
${\rm cost}(f^\star) = {\rm cost}(f) + {\rm cost}(P)$.
\end{theorem}
\begin{proof}
Either $f(a)=1$ or $f(a)=0$. In case $f(a)=1$, $f^\star(a)=1$, 
$P = d, x_i$, ${\rm cost}(P) = 0$ and we are done. 
In case $f(a)=0$, first note that there exists a $d-x_i$ path in
$G_f$. Namely, there is exactly one $d^\prime \in D_i$ for which 
$f((x_i, d^\prime)) = 1$,
which allows the path $d, t, d^\prime, x_i$. Let $P$ be a 
minimum-cost $d-x_i$ path in $G_f$. Together with arc $(x_i, d)$
$P$ forms a circuit $C$. The directed circuit $C$ in $G_f$ gives an
undirected circuit in $G^a$. For all $b \in A$, define flow $f^\star$
in $G^a$ as follows:
\begin{displaymath}
  f^{\star}(b) =
  \left\{
    \begin{array}{ll}
      0 & {\rm if} \; b^{-1} \in C\\
      1 & {\rm if} \; b \in C\\
      f(b) & {\rm else.}
    \end{array}
  \right.
\end{displaymath}
It is easy to check that $f^\star$ is again a flow of maximum value.

Because $f$ is extreme, we may assume that $P$ enters and leaves $t$
only once, say via arcs $b_{\rm in}$ and $b_{\rm out}$ respectively
(where $b_{\rm in} = (d, t)$). It follows that ${\rm cost}(P) = 
w(b_{\rm in}) - w(b_{\rm out})$. From Theorem~\ref{thm:circuit} we know
that ${\rm cost}(P) \geq 0$. Similarly,
\begin{displaymath}
\begin{array}{rl}
{\rm cost}(f^\star) & = \sum_{b \in A} f^\star(b) w(b) \\
 & = \sum_{b \in A} f(b) w(b) + w(b_{\rm in}) - w(b_{\rm out}) \\
 & = {\rm cost}(f) + {\rm cost}(P)
\end{array}
\end{displaymath}

It remains to show that $f^\star$ is extreme in $G^a$. Suppose not,
i.e. there exists a flow $g$ in $G^a$ with maximum value and
${\rm cost}(g) < {\rm cost}(f^\star)$. As ${\rm cost}(f^\star) = 
{\rm cost}(f) + {\rm cost}(P)$ and ${\rm cost}(P) \geq 0$, there are
two possibilities. The first is that ${\rm cost}(g) < {\rm cost}(f)$,
which is not possible because $f$ is extreme. The second is that there
exists an $x_i-d$ path $P^\prime$ in $G_f$ with 
${\rm cost}(P^\prime) < {\rm cost}(P)$ which also leads to a
contradiction because $P$ is a minimum-cost path. Hence $f^\star$ is
extreme.
\end{proof}

\begin{algorithm}[t]
\caption{More efficient hyper-arc consistency}
\label{alg:hac2}
\begin{algorithmic}
  \STATE{compute minimum-cost flow $f$ in $G$}
  \IF{cost$(f) > \max{D_z}$}
  \STATE{return {\sc Inconsistent}}
  \ENDIF
  \IF{$\min{D_z} < {\rm cost}(f)$}
  \STATE{set $\min{D_z} = {\rm cost}(f)$}
  \ENDIF
  \FOR{$a = (x_i, d)$ with $f(a) = 0$}
  \STATE{compute minimum-cost $d - x_i$ path $P$ in $G_f$} 
  \IF{cost$(f)$ + cost$(P)$ $>$ $\max{D_z}$} 
  \STATE{remove $d$ from $D_i$}
  \ENDIF
  \ENDFOR
\end{algorithmic}
\end{algorithm}

Algorithm~\ref{alg:hac2} first computes a minimum-cost flow $f$ in
$G$. This takes $O(nm)$ time, as we have seen in
Section~\ref{ssc:softflow}.
If ${\rm cost}(f) > \max{D_z}$, we know that the \softad\
constraint is inconsistent. If this is not the case, we update 
$\min{D_z}$. Next, we scan all arcs $a = (x_i, d)$ for which 
$f(a) = 0$. For each of these arcs, we compute a minimum-cost $d-x_i$ 
path $P$ in $G_f$. By Theorem~\ref{thm:path} and Theorem~\ref{thm:hac}, 
we remove $d$ from $D_i$ if cost$(f)$ + cost$(P)$ $>$ $\max{D_z}$.
This can be done efficiently, as shown by the following theorem.
\begin{theorem}\label{thm:filter}
Let $\textup{\softad}(x_1, \dots, x_n, z)$ be consistent and $f$ an
integer minimum-cost flow in $G$. Then 
$\textup{\softad}(x_1, \dots, x_n, z)$ can be made hyper-arc consistent 
in $O(m)$ time.
\end{theorem}
\begin{proof}
The complexity of the filtering algorithm depends on the computation of 
the minimum-cost $d-x_i$ paths for arcs $(x_i, d)$. We make use of the
fact that only arcs $a \in A_t$ contribute to the cost of such a path.

Consider the strongly connected components\footnote{A strongly connected
component in a directed graph $G = (V, A)$ is a subset of vertices 
$S \subseteq V$ such that there exists a directed $u-v$ path in $G$ for 
all $u, v \in S$.}
of the graph $\tilde{G}_f$ which is a copy of $G_f$ where $s$ and $t$ 
and all their incident arcs are removed.
Let $P$ be a minimum-cost $d-x_i$ path $P$ in $G_f$.
If $P$ is equal to $d,x_i$ then $f(x_i, d) = 1$ and ${\rm cost}(P) = 0$.
Otherwise, either $x_i$ and $d$ are in the same strongly connected 
component of $\tilde{G}_f$, or not. In case they are in the same strongly 
connected component, $P$ can avoid $t$ in $G_f$, and ${\rm cost}(P) = 0$. 
In case $x_i$ and $d$ are in different strongly connected components of 
$\tilde{G}_f$, say $x_i \in S_1$ and $d \in S_2$, we have
\begin{equation}\label{eq:minmax}
{\rm cost}(P) = \min_{
\begin{scriptsize}
\begin{array}{c}
  a \in \{ (d^\prime, t) \mid (d^\prime, t) \in A_f,\\ d^\prime \in S_2\}
\end{array} \end{scriptsize}} w(a)
+ \min_{
\begin{scriptsize}
\begin{array}{c}
  a \in \{ (t, d^{\prime\prime}) \mid (t, d^{\prime\prime}) \in A_f,\\
d^{\prime\prime} \in S_1 \textrm{ or } (d^{\prime\prime},x_i) \in A_f \}
\end{array} \end{scriptsize}} w(a).
\end{equation}
Property (\ref{eq:minmax}) follows from the fact that $P$ uses exactly
one ingoing and one outgoing arc for $t$.

Arcs $a$ with $f(a) = 1$ or within a strongly connected component will 
all use a minimum-cost path with cost equal to 0, and will therefore be 
all consistent if cost$(f) \leq \max{D_z}$.
For all other arcs, we can resort to property (\ref{eq:minmax}). For 
this we only have to compute once for each strongly connected component 
$S$ of $\tilde{G}_f$ the minimum-cost arc going from $S$ to $t$ and the
minimum-cost arc going from $t$ to $S$ (if such arcs exist), which takes 
in total $O(m)$ time.
The strongly connected components of $\tilde{G}_f$ can be computed in 
$O(n+m)$ time, following Tarjan \cite{tarjan72}. Hence the total time 
complexity of achieving hyper-arc consistency is $O(m)$, as $n < m$.
\end{proof}

The proof of Theorem~\ref{thm:filter} applies to any constraint whose 
graph representation resembles $G$ and has only costs on arcs from 
$D_X$ to $t$. For all such constraints that are consistent, hyper-arc 
consistency can be achieved in $O(m)$ time. Note that this is equal 
to the complexity of achieving hyper-arc consistency on these 
constraints if no costs are involved.

\section{Conclusion and Discussion}\label{sc:concl}
We have presented an algorithm that checks consistency of the \softad\ 
constraint on $n$ variables in $O(nm)$ time and achieves hyper-arc 
consistency in $O(m)$ time, where $m$ is the sum of the cardinalities 
of the domains. 
A previous method for removing domain values that are inconsistent with 
the \softad\ constraint did not ensure hyper-arc consistency \cite{petit01}.
Moreover, that method has a time complexity of $O(m^2 n \sqrt{n})$. 
Hence our algorithm improves on this in terms of quality as well 
as time complexity.

The \softad\ constraint is related to the standard \alld\
constraint \cite{regin} and the minimum weight \alld\ constraint
\cite{caseau_laburthe_cp97}. The minimum weight \alld\ constraint is a 
particular instance of the global cardinality constraint with costs
\cite{regin99,regin2002}.
For that constraint, hyper-arc consistency can be achieved in 
$O(n(m + d \log d))$ time, where $d$ is the cardinality of the union 
of all domains \cite{regin99,regin2002,sellmann02}. It is achieved by 
finding $n$ shortest paths, each taking $O(m + d \log d)$ time to compute.
Although our algorithm has a similar flavour, the underlying graphs 
have a different cost structure. We improve the efficiency by 
exploiting the cost structure of our particular graph when computing 
the shortest paths. Our result can be applied to other constraints 
with a similar graph representation and cost structure.

\section*{Acknowledgements}
Many thanks to Bert Gerards for valuable comments. Thanks also go to 
Sebastian Brand for fruitful discussion. Finally, the constructive 
remarks of Jean-Charles R\'egin were highly appreciated.

\end{document}